# Cross-domain Self-supervised Framework for Photoacoustic Computed Tomography Image Reconstruction

Hengrong Lan, Lijie Huang, Zhiqiang Li, Jing Lv, and Jianwen Luo, *Senior Member, IEEE*

*Abstract*—Accurate image reconstruction is crucial for photoacoustic (PA) computed tomography (PACT). Recently, deep learning has been used to reconstruct PA image with a supervised scheme, which requires high-quality images as ground truth labels. However, practical implementations encounter inevitable trade-offs between cost and performance due to the expensive nature of employing additional channels for accessing more measurements. Here, we propose a cross-domain self-supervised (CDSS) reconstruction strategy to overcome the lack of ground truth labels from limited PA measurements. We implement the self-supervised reconstruction in a model-based form. Simultaneously, we take advantage of self-supervision to enforce the consistency of measurements and images across three partitions of the measured PA data, achieved by randomly masking different channels. Our findings indicate that dynamically masking a substantial proportion of channels, such as 80%, yields meaningful self-supervisors in both the image and signal domains. Consequently, this approach reduces the multiplicity of pseudo solutions and enables efficient image reconstruction using fewer PA measurements, ultimately minimizing reconstruction error. Experimental results on in-vivo PACT dataset of mice demonstrate the potential of our self-supervised framework. Moreover, our method exhibits impressive performance, achieving a structural similarity index (SSIM) of 0.87 in an extreme sparse case utilizing only 13 channels, which outperforms the performance of the supervised scheme with 16 channels (0.77 SSIM). Adding to its advantages, our method can be seamlessly deployed on different trainable models in an end-to-end manner, further enhancing its versatility and applicability.

## I. INTRODUCTION

PHOTOACOUSTIC imaging (PAI) is a hybrid imaging modality that combines the advantages of optical imaging with an acoustic detection. It has found extensive applications in clinical research and translation due to its high imaging depth-to-spatial resolution ratio (>100) [1-5]. Meanwhile, rich contrasts can be achieved owing to the optical excitation [6-8]. Photoacoustic (PA) computed tomography (PACT) is a prominent approach among different implementations of PAI. PACT involves illuminating biological tissues with non-focused ultra-short pulses and capturing photoacoustic signals using an array of ultrasound transducers from multiple angles. This technique allows for high-speed, deep-tissue imaging with functional contrasts, making it suitable for clinical and translational applications [9-16]. A key aspect of influencing the quality of PACT is the sophisticated reconstruction of the initial pressure distribution, also known as the PA image, based on ultrasound time-of-flight (TOF) information [17].

In practice, the quality of PACT image is also influenced by the number of detection channels and the angle of views. Insufficient channels or limited views can lead to image artifacts and blurring [18, 19]. These challenges give rise to an ill-posed problem in PACT image reconstruction, which in turn affects image quality and resolution. To address this, iterative model-based algorithms have been employed, incorporating various regularization techniques [20], e.g., total variation (TV) [21], wavelet sparsity [22], L1 sparsity [23], and Tikhonov [24]. However, model-based methods face a trade-off between iterative time consumption and image quality. Additionally, the imposed regularization can inadvertently lead to information loss or introduce errors into the reconstructed features.

In recent years, deep learning (DL) has emerged as a valuable approach for reconstructing biomedical images using data-driven methods [25, 26]. In the field of PACT, DL-based methods have been successfully applied to image reconstruction [27-29]. Initially, convolutional neural networks (CNNs) are utilized to localize point-like targets and remove reflection artifacts from PA images or pre-beamformed data [30, 31]. Another straightforward approach involves directly reconstructing PA signals into images using end-to-end CNN

This study was supported in part by the National Natural Science Foundation of China (NSFC) (No. 62027901). (Corresponding author: Jianwen Luo)

Hengrong Lan, Lijie Huang, Zhiqiang Li, and Jianwen Luo are with the Department of Biomedical Engineering, School of Medicine, Tsinghua University, Beijing 100084, China (e-mail: luo_jianwen@tsinghua.edu.cn).

Jing Lv is with the Research Center of Medical Sciences, Guangdong Provincial People's Hospital, Guangdong Academy of Medical Sciences, Guangzhou 510000, China.

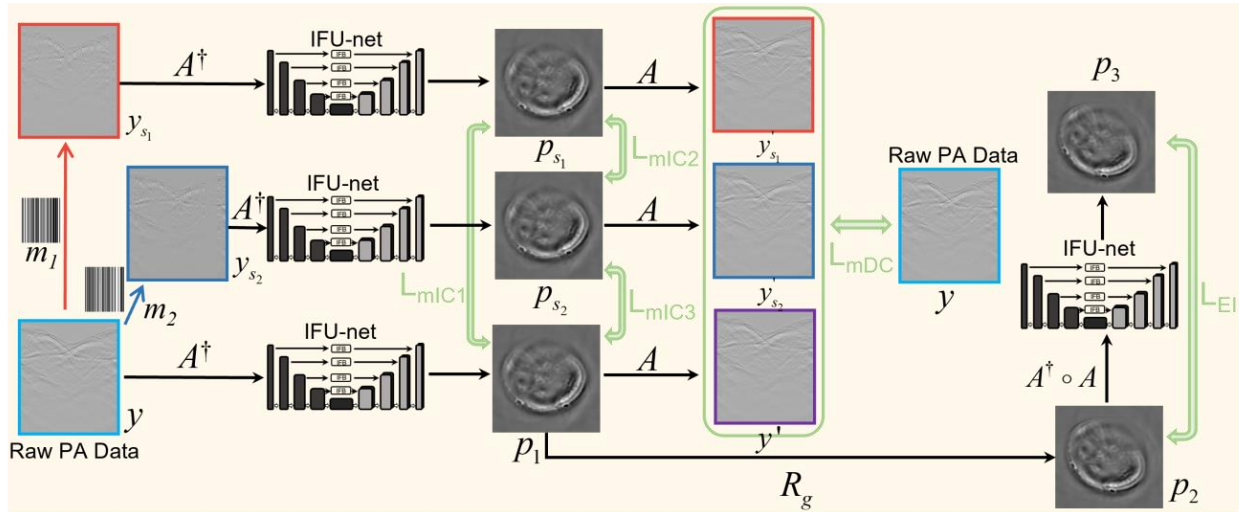

Fig. 1. The illustration of cross-domain self-supervised (CDSS) framework. In training procedure, two disjointed random channel masks ($m_1$ and $m_2$) are used to separate the raw data y into $y_{s1}$ and $y_{s2}$. These data (y, ys1 and ys2) are then fed into $A^†$ and IFUnet to reconstruct three images ($p_1$, $p_{s1}$ and $p_{s2}$). The outputs are restricted in signal (masked data consistency loss) and image (masked image consistency loss) domains. In addition, full sampled image $p_1$ is transformed to $p_2$ by a rotated transformation $R_g$ and passes it to $A^†∘A$ and IFUnet to produce $p_3$, which can be enforced by equivariance loss. In evaluating procedure, the trained model can process the raw data with/without masking directly.

architectures [32-34]. These methods are effective and straightforward for reconstructing simple targets. DL models can also be employed for PA signal processing, such as enhancing the signal-to-noise ratio (SNR) or expanding the bandwidth of PA signals [35, 36]. Improved PA signals can then be reconstructed with conventional reconstruction methods to obtain high-quality images. Furthermore, the performance of CNNs can be significantly enhanced if the input PA signals are processed as pre-reconstructed data associated with target locations [37, 38].

In the case of complex targets, image-to-image strategies have been proved to be effective in addressing them as an image processing task. CNNs can be employed to remove artifacts in PA images [39-43]. In scenarios with limited views or sparse data acquisition, missing information in the image can be supplemented using additional data. DL models have been shown to be promising in improving the SNR of PA images affected by factors like low laser pulse energy and amplifier performance [44-46]. Furthermore, by combining PA images with PA signals, CNNs can further enhance the quality of PA images compared to using a single input modality [47-49]. In situations where numerous unpaired data are available (unpaired low-quality and high-quality images), the semi-supervised generative adversarial networks (GANs) can generate pseudo-labels to enhance the quality of PA images [50].

DL techniques have also been employed to enhance model-based methods by introducing learnable components, thereby improving quality in the case of reducing computational cost. On one hand, CNNs can be utilized to learn the regularization terms, allowing for adaptive learning of priors from data [51-54]. On the other hand, a learnable iterative step based on a compact CNN architecture can be developed to integrate the entire iteration process [55, 56]. Moreover, CNNs have been utilized for joint tasks, simultaneously improving results across different objectives[57, 58]. The powerful capabilities of untrained CNNs can also be harnessed to fit the PA signal within a model-based reconstruction framework [59].

However, these approaches typically rely on high-quality images as ground truth labels for learning the mapping, which limits their applicability as acquiring such high-quality images can be expensive or even impossible in real-world scenarios. Consequently, breaking free from this limitation and achieving effective reconstruction independent of high-quality ground truth labels has become a meaningful pursuit in the field of medical image reconstruction [60-63].

In this work, we propose an unsupervised strategy for PACT reconstruction, aiming to achieve stable reconstruction from incomplete measurement data. Our approach, called cross-domain unsupervised reconstruction (CDSS), exploits the limited channels and the physical properties of PACT, such as equivariance and sparsity, to learn beyond the compressed data space (see Fig. 1). We train CDSS using a limited dataset by randomly splitting 128 channels into two complementary sets with dynamic masking. Both the image and signal domains are enforced through combined objective functions that involve the original incomplete data and masked data. Moreover, an image-to-image CNN architecture (IncepFormer Unet, IFUnet) is proposed to reconstruct the PA image in CDSS. Notably, this work represents an instance of achieving unsupervised learning in PA imaging without relying on higher-quality data as ground truth labels, which offers a unique advantage in scenarios where high-quality data may not be readily available.

We validate the superiority of CDSS using an *in-vivo* mice dataset. Experimental results demonstrate that our approach maintains good performance even when the number of channels decreases from 128 to 13. Furthermore, we compare CDSS with a supervised method using different numbers of channels. The comparison reveals that CDSS outperforms the supervised method, particularly in extreme cases with fewer than 16 channels. Our major contributions can be summarized as follows:

1. We proposed a self-supervised learning scheme for PACT reconstruction, which eliminates the need for

high-quality labels. This innovative approach leverages cross-domain self-supervision, sparse regularization, as well as the equivariance property of PACT, and offers a possibility in cases where high-quality data cannot be accessed.

2. We have proven the effectiveness of a random masking strategy on measurement channels. This strategy significantly enhances the quality of reconstructed images, especially in extremely sparse scenarios with fewer than 16 channels, marking a significant improvement in image reconstruction quality.

3. We have proposed a hybrid architecture for PACT image reconstruction. This architecture integrates an Inception Mixer module, which combines the strengths of convolutional layers and max-pooling, with Transformers. This architecture could be used for other image-to-image enhancement.

## II. METHODS

### A. Self-supervision for PACT image reconstruction

In PACT, the measurement data $y$ can be modeled as an acoustic forward problem. The linear forward operator $A$ maps the initial pressure distribution $p$ (PA image) to the measurement data $y$:

$$y = Ap + \varepsilon, \quad (1)$$

where $\varepsilon$ is the additive noise. To reconstruct the initial pressure distribution $p$ from the given measurement data $y$, we solve a least-squares minimization problem with a regularization term $R(p)$:

$$\min_p \frac{1}{2}\|Ap - y\|_2^2 + \lambda R(p), \quad (2)$$

where $\lambda > 0$ is the parameter to balance the proportion of the first term (data consistency) and the second term (prior information about $p$). The regularization term $R(p)$ is introduced to impose additional constraints or prior knowledge on the reconstructed pressure distribution, promoting certain desired properties such as smoothness or sparsity. The specific form of the regularization term depends on the characteristics of the imaging problem and the desired properties of the reconstructed image. By solving this optimization problem, we aim to find the pressure distribution $p$ that best fits the measurement data $y$ while satisfying the regularization, leading to an accurate and high-quality reconstruction in PACT.

In this work, we propose CDSS for training a reconstruction model $M$ using a self-supervised approach, which is embodied in a similar objective as the optimization problem in Eq. (2):

$$\min_\Theta \frac{1}{2}\|AM_\Theta(A^\dagger y) - y\|_2^2, \quad (3)$$

where $M$ is a deep learning model with trainable weights $\Theta$ (i.e., IFUnet in this work) and $A^\dagger$ is the pseudo-inverse operator. The objective of CDSS is to minimize the discrepancy between the reconstructed PA image and the measurement data by Eq. (4), while incorporating additional constraints or prior knowledge to promote desirable properties in the reconstructed image. This objective is formulated in a self-supervised manner, where the supervision signal is derived from the data itself, rather than relying on external ground truth images.

CDSS incorporates three key designs to improve image quality with incomplete measurements, as illustrated in Fig. 1. During the training stage, the measurement data $y$ is randomly split into two complementary subsets. These two subsets are then used as inputs for reconstructing the image using the operator $A^\dagger$ and IFUnet. The errors between the reconstructed images ($p_{s1}$, $p_{s2}$, and $p_1$) and the inferred measurement data ($y_{s1}'$, $y_{s2}'$, and $y'$) are computed, respectively. Furthermore, the entire image $p_1$ is transformed to $p_2$ using a specific transformation represented by $R_g$, and then converted to the inferred measurement data using the operator $A$. The inferred measurement data is then fed into $f$, where $f$ is a function defined as $f = M \circ A^\dagger$ (where $M$ represents IFUnet), to reconstruct $p_3$. Once trained, $f$ can directly reconstruct the PA image $p$ using masked or unmasked measurements for evaluation. Additional details will be described in the following sections.

### B. Masked consistency in cross-domain

During the training stage, we partition the entire measurement data $y$ into two subsets along the channel dimension. We randomly choose a given proportion of channels from all the channels, and each channel is equally likely to be selected in this procedure. In each batch, we generate a mask $m_1$ with a specified sampling ratio (referred to as the masking ratio). Additionally, we compute the complementary mask $m_2$ by subtracting $m_1$ from 1 ($m_2 = 1 - m_1$). By performing element-wise multiplication with $m_1$ and $m_2$, we obtain two subsets of the measurement data: $y_{s1}$ and $y_{s2}$, respectively:

$$y_{s1} = m_1 \odot y, \quad y_{s2} = m_2 \odot y. \quad (4)$$

The masks $m_1$ and $m_2$ are randomly generated for each batch, ensuring a random distribution of valid channels. For example, if $m_1$ has a masking ratio of 20%, $y_{s1}$ will have 80% randomly selected valid channels, while $y_{s2}$ will have 20% valid channels for each batch. This strategy allows our approach to handle specific scenarios such as limited or sparse views. Subsequently, $y_{s1}$ and $y_{s2}$ are used as inputs to the reconstruction function $f$, producing the reconstructed images $p_{s1}$ and $p_{s2}$, respectively. These reconstructed images should be consistent with the image $p_1$, which is reconstructed from the original measurement data $y$.

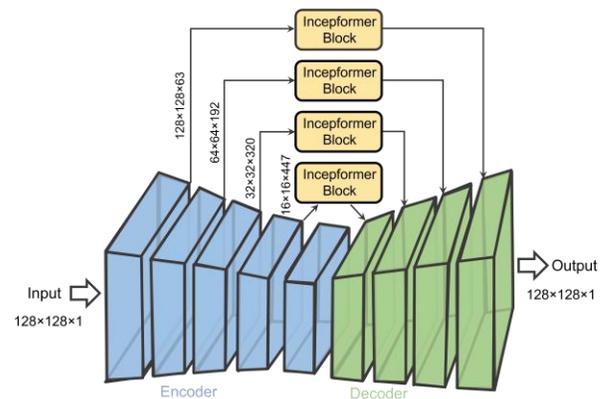

Fig. 2. The overall architecture of IFUnet.

To enforce this consistency, we compute a masked image consistency (mIC) loss using the $L_1$ norm:

$$L_{mIC} = \|p_{s1} - p_{s2}\|_1 + \|p_1 - p_{s1}\|_1 + \|p_1 - p_{s2}\|_1. \quad (5)$$

The dynamic masking strategy allows us to capture information from all the 128 channels during training by utilizing sub-sampling without concerns about information loss caused by sparse input. In order to achieve better unsupervised performance, we also enforce data consistency in the signal domain, similar to the approach used in model-based methods. We introduce a masked data consistency (mDC) loss defined as follows:

$$L_{mDC} = \|y'_{s1} - y\|_2^2 + \|y'_{s2} - y\|_2^2 + \|y' - y\|_2^2, \quad (6)$$

where $y_{s1}'$, $y_{s2}'$, and $y'$ are the inferential measurement data and can be written as:

$$y'_{s1} = Ap_{s1}, y'_{s2} = Ap_{s2}, y' = Ap_1. \quad (7)$$

Instead of enforcing each other, the ground truth $y$ serves as a uniform constraint in Eq. (6). In summary, the model is trained using three sets. The high-ratio mask eliminates most channels, creating a challenging task of reconstructing the image using the remaining few channels. Conversely, the complementary mask creates a contrasting task where the image can be more easily recovered using relatively large number of channels.

### C. Equivariance for PACT

Since the number of detection channels is smaller than the dimension of the image space, the operator $A$ has a non-trivial null space. We utilize the invariance of PACT for rotated transformations as an additional prior, where the same tissue can be formed at any angles. For an image $p$ from a set of PA images $P$ and a unitary matrix $R_g$ with arbitrary rotation $g$, due to the invariant property, we have:

$$R_g p \in P. \quad (8)$$

Therefore, we obtain an equivariance of the transformation $R_g$ with $M \circ A^\dagger \circ A$:

$$f(AR_g p) = R_g f(Ap). \quad (9)$$

Based on the above ideas, we obtain $p_2$ after rotating $p_1$ by an arbitrary angle ($p_2 = R_g p_1$) and obtain $p_3$ after $A$ and $f$ ($p_3 = MA^\dagger Ap_2$). Namely, we can compute equivariant imaging (EI) loss to impose equivariance as shown in Fig. 1:

$$L_{EI} = \|p_2 - p_3\|_2^2. \quad (10)$$

Ref. [63] proved that this EI constraint allows us to learn beyond the limited measurement $y$. Noting that the following combined matrix $O$ should be as big as possible with a full rank:

$$O = \begin{bmatrix} AR_g^1 & AR_g^2 & \cdots & AR_g^n \end{bmatrix}^T. \quad (11)$$

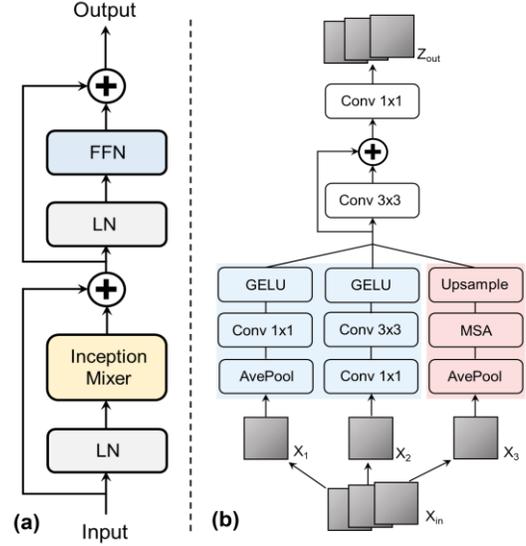

Fig. 3. (a) The details of incepformer block. (b) The details of Inception Mixer [66].

### D. Inception transformer U-net

Recently, vision transformer (ViT) has achieved excellent performance in many fields of computer visions [64]. In the CDSS framework, we introduce a hybrid architecture called IFUnet, which combines CNN and Transformer modules. The overall architecture is illustrated in Fig. 2. We use Unet as the backbone [65], and hybrid modules of CNN and ViT (referred as Incepformer blocks) are employed to process features of different sizes from the encoder to the decoder of Unet, as shown in Fig. 3 (a). In Fig. 3 (a), LN represents the LayerNorm layer, and FFN represents the feed-forward network.

The Incepformer block is a transformer module, with the addition of the Inception Mixer, which combines multi-head self-attention (MSA) with convolution operations [66]. This structure resembles the parallel structure of the inception module, as depicted in Fig. 3 (b). This combination allows the model to capture both local and global representations in the skipped connections. The input features are divided into three parts along the channels, and the relationships within each part are calculated separately:

$$Z_1 = GELU(Conv(AvePool(X_1))),$$
$$Z_2 = GELU(Conv_{3\times 3}(Conv_{1\times 1}(X_2))), \quad (12)$$
$$Z_3 = Upsample(MSA(AvePool(X_3))),$$

where GELU indicates Gaussian Error Linear Unit, which is an activation function. Then we compute the fusion of three portions. The output features can be expressed as:

$$Z_{cat} = Concat(Z_1, Z_2, Z_3),$$
$$Z_{out} = Conv_{1\times 1}(Z_{cat} + Conv_{3\times 3}(Z_{cat})). \quad (13)$$

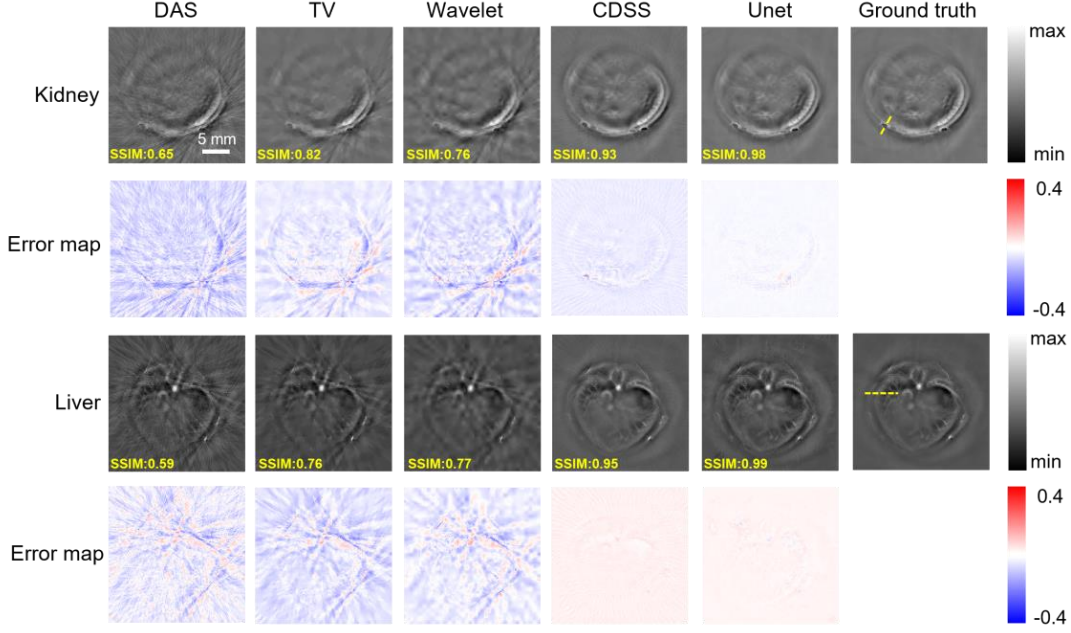

Fig. 4. The experimental results of in-vivo mice using 64 channels. DAS: delay-and-sum; TV: total variation; CDSS: our self-supervised method; Unet: supervised Unet; Ground truth reconstructed by DAS from 512 channels. The error map indicates the difference between the corresponding image and ground truth.

### E. Sparse regularization in image domain

To enforce the sparsity of the reconstructed images in the wavelet domain, we use the $L_1$ norm as a measure of sparsity [20]. This regularization term encourages a sparse representation of the image coefficients in the wavelet transform domain. By minimizing the $L_1$ norm, we promote a solution that contains fewer non-zero coefficients, leading to a sparser representation and potentially better reconstruction quality:

$$L_{DWT} = \|\psi p_1\|_1 + \|\psi p_{s1}\|_1 + \|\psi p_{s2}\|_1, \quad (14)$$

where $\Psi$ is the forward discrete wavelet transform. To promote smoothness in the output images, we employ TV regularization, which encourages the reconstructed images to have sharp boundary features while suppressing small variations or noise by taking advantage of numerical differentiation to penalize large gradients in the image:

$$L_{TV} = \|\nabla p_1\|_1 + \|\nabla p_{s1}\|_1 + \|\nabla p_{s2}\|_1. \quad (15)$$

By minimizing the TV regularization term, we encourage the reconstructed images to have smooth transitions and well-defined boundaries, resulting in visually pleasing and coherent outputs.

These priors constrain the output of model in the image domain. Finally, we train CDSS with a combination of the above loss functions:

$$L_{final} = \lambda_{mDC} L_{mDC} + \lambda_{mIC} L_{mIC} + \lambda_{EI} L_{EI} + \lambda_{DWT} L_{DWT} + \lambda_{TV} L_{TV}, \quad (16)$$

where $\lambda_{mDC}$, $\lambda_{mIC}$, $\lambda_{EI}$, $\lambda_{DWT}$ and $\lambda_{TV}$ embody the proportion of different regularizations.

## III. EXPERIMENTS

### A. Dataset preparation and implementation

We acquired *in-vivo* mice data with a panoramic PACT system (SIP-PACT-512, Union Photoacoustic Technologies Co., Ltd., Wuhan, China). The system consists of an optical parametric oscillator (OPO) laser with a pulse repetition frequency of 10 Hz. We used a 1064 nm wavelength for illumination, which provides high-quality structural imaging with minimal scattering. For all the datasets, PA signals were generated and received using a 360° ring-shaped ultrasound transducer array with 512 channels (radius: 50 mm, central frequency: 5 MHz). The acquired PA data had 2000 samples with 40 MHz sampling rate.

In experiments, a total of four healthy nude mice (8-week-old, SPF Biotechnology Co., Ltd., Beijing, China) were used for imaging. The mice were placed in a temperature-controlled water tank during the imaging process. We performed whole-body scans with a step size of 0.02 mm by moving the animal holder using a positioner. All experimental procedures were approved by the Institutional Animal Care and Use Committee in Guangdong Provincial People's Hospital. To ensure the

TABLE I.
QUANTITATIVE COMPARISONS OF DIFFERENT METHODS WITH DIFFERENT NUMBERS OF CHANNELS (MEAN ± STANDARD DEVIATION)

| 128 | SSIM ↑ | PSNR ↑ | RMSE ↓ | 64 | SSIM ↑ | PSNR ↑ | RMSE ↓ |
|---|---|---|---|---|---|---|---|
| DAS | 0.773±0.048 | 24.887±2.280 | 0.038±0.007 | DAS | 0.579±0.068 | 23.961±3.019 | 0.062±0.024 |
| TV | 0.898±0.058 | 30.407±3.338 | **0.032±0.012** | TV | 0.769±0.088 | 26.788±3.789 | 0.051±0.025 |
| Wavelet | 0.821±0.040 | 27.748±3.565 | 0.045±0.021 | Wavelet | 0.802±0.056 | 26.535±4.225 | 0.053±0.028 |
| CDSS | **0.917±0.049** | **32.422±6.591** | 0.033±0.033 | CDSS | **0.917±0.049** | **32.422±6.591** | **0.033±0.033** |
| Unet | 0.978±0.010 | 38.758±4.534 | 0.013±0.008 | Unet | 0.953±0.022 | 31.008±6.025 | 0.028±0.019 |

Unet is the supervised scheme using the DAS-reconstructed image from 512 channels as ground truth.
* Higher SSIM values, higher PSNR values and smaller RMSE values indicate higher performance.

fairness of the experiments, we selected three mice for the training set, while the remaining mouse was used as the test set. After removing some of the more intrusive faults, the training dataset was comprised of 3000 slices, while the test dataset consisted of 400 slices.

The framework is implemented in Pytorch [67]. The discrete forward model $A$ is established in MATLAB (The Mathworks, Inc., Natick, MA, USA), which is referred to a curve-driven method [68]. Specifically, we operate under the assumption that the acoustic properties are homogeneous, and we set the speed of sound at 1513 m/s for forward and inverse processes. The implementation environment is composed of an Intel Xeon E5-2620 CPU with 128 GB RAM and four NVIDIA Titan V GPUs with 12 GB memory. The AdamW optimal method [69] is used to train our approach. In the training stage, the batch size is 32 for all models, and the total iteration is 400 epochs with a 0.001 initial learning rate. The parameters of Eq. (16) ($\lambda_{mDC}$, $\lambda_{mIC}$, $\lambda_{EI}$, $\lambda_{DWT}$ and $\lambda_{TV}$) are 3, 13, 6, 0.002 and 0.001 respectively.

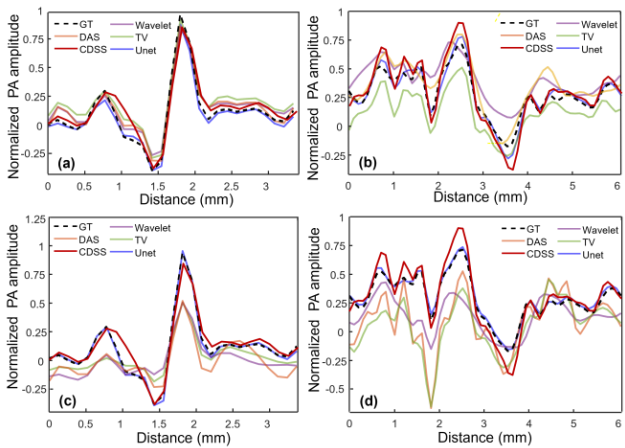

Fig. 5. Comparison of the profiles extracted from the yellow dashed lines in the results with different numbers of channels in Fig. 4. The profiles of (a) kidney and (b) liver reconstructed with 128 channels. The profiles of (c) kidney and (d) liver reconstructed with 64 channels. GT: ground truth; DAS: delay-and-sum; CDSS: our unsupervised method; TV: total variation; Unet: supervised Unet.

*B. Comparison with other methods*

To evaluate the performance of CDSS, we compared it with several benchmark methods, including wavelet sparsity, TV, and a direct reconstruction method using delay-and-sum (DAS). Given that DAS is the most widely employed non-iterative reconstruction method, and that the input of the IFUnet is reconstructed by $A^\dagger$, which is based on DAS, this comparison provides intuitive conclusions. Therefore, contrasting our method with DAS offers valuable insights into the effectiveness of IFUnet. Regarding the selection of hyperparameter for iterative reconstructions, we conducted a search within a logarithmically spaced range spanning from 0.0001 to 0.05. This exploration enabled us to identify the optimal parameters that yield superior results. Finally, the denoising weight of wavelet is 0.0002 and the denoising weight of TV is 0.0005. Concurrently, the total steps are 500 for all iterative method. For enhancing the convergence speed of the iterative method, we initialized the iteration process with a relatively large step size (0.5). Subsequently, we iteratively reduced the step size by half after every 100 iterations. We also compared CDSS with the supervised Unet method [65]. The batch size is 32, and the total iteration is 400 epochs with a 0.001 initial learning rate.

Fig. 4 shows the results of two slices from the test data, one from the kidney and another from the liver, using 64 channels. The term "ground truth" of all the results below pertains to the reconstructed images derived from the acquired 512-channel data using DAS. It is important to note that for DL methods, **CDSS only has access to the measurement data with 128 channels**, while supervised Unet is trained using high-quality images reconstructed from 512 channels using DAS. The masking ratio used for the comparisons is 50%, meaning that the same model can reconstruct the image from both 128 and 64 channels.

When using only 64 channels, the results obtained by the DAS method in Fig. 4 show severe artifacts that hinder the detailed structure of the mouse, except for the outline of the abdomen and strong absorbers like the spleen in the kidney or the inferior vena cava in the liver. The error map also reflects the presence of global artifacts that affect the representation of details. The TV and wavelet sparsity regularizations partially suppress these artifacts, resulting in more concentrated errors within the body, as shown in the error map. The CDSS method, on the other hand, shows complete information of the mouse viscera with minor artifacts. This indicates that CDSS performs similarly to the supervised Unet method. However, slightly blurry at the edges of the body appears in the CDSS results, suggesting that similar high-frequency information might have been removed when removing the artifacts using CDSS. In the results, we further mark the value of structural similarity index (SSIM) for each result. For a fair comparison, we also evaluated different methods using 128 channels, as shown in Fig. S1 in the supplementary materials.

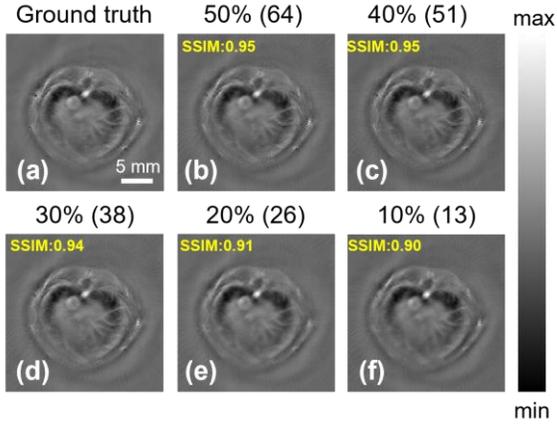

Fig. 6. The results of CDSS with different masking ratios. (a) Ground truth reconstructed by DAS from 512 channels. The whole 128 channels are randomly masked with different ratios. The image reconstructed by CDSS from random (b) 64, (c) 51, (d) 38, (e) 26, and (f) 13 channels.

In Fig. 5, we present profiles along the yellow dashed lines in Fig. 4 for both 64 and 128 channels, comparing the performances of different methods. With 128 channels, most methods demonstrate good performance compared to the ground truth, disregarding stripe artifacts in the background. The results of the DL methods exhibit consistent spatial resolution as shown in Fig. 5 (a) and (b). Conventional methods (DAS, TV, and wavelet sparsity) show more interference in the liver profiles (Fig. 5 (b) and (d)). Notably, Fig. 5 (d) illustrates that the profiles of these methods exhibit significant variations when the number of channels decreases to 64. CDSS produces smoother results, although with some artifacts compared to other methods (Fig. 5 (c)).

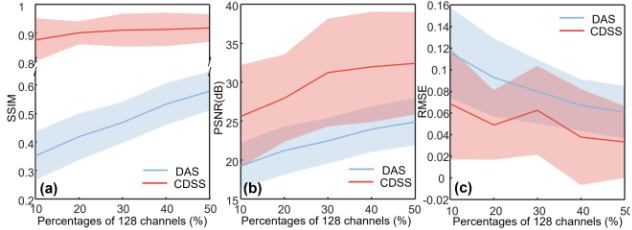

Fig. 7. The (a) SSIM, (b) PSNR, and (c) RMSE of CDSS and DAS when the number of masked channels changes from 90% to 50% of 128.

The quality of the reconstructed images obtained by different methods is evaluated using SSIM, peak signal-to-noise ratio (PSNR), and root mean squared error (RMSE). The quantitative results for the entire test set, both with 128 and 64 channels, are presented in Table I. With 128 channels, all the methods demonstrate good performance. DAS has the lowest SSIM due to artifacts, while images with fewer artifacts exhibit more texture information compared to the ground truth. Except for supervised Unet, CDSS consistently outperforms the other methods in terms of SSIM, PSNR, and RMSE, demonstrating superior performance even with minor artifacts. Even in comparison to the iterative TV method, CDSS demonstrates a relatively comparable performance when evaluated on the 128-channel data. It is noteworthy that our method achieves these results without requiring iterative procedures, which underscores its meaningful contribution. Additionally, CDSS maintains its high performance even when the number of channels is reduced to 64, further highlighting its effectiveness in the situation of few channels compared with iterative methods.

### C. Masking strategy for measurement channel

The random masking strategy can decrease the number of channels by giving different ratios. That is employed to train a single model simultaneously using three different numbers of channels, resulting in consistent performance across these inputs. Simultaneously, the varying masking introduces diverse down-sampling methods, making the proposed method adept at handling different artifacts effectively. To examine the impact of masking ratio, we vary the ratio to assess its influence of the number of valid channels from 10% to 50%. In this section, we focus solely on exploring the difference between the input and output of the model of CDSS as the masked channel varies. We achieve this by drawing comparisons with the direct reconstruction method, DAS. This comparison allows us to understand whether our method can sustain high performance when the input image is progressively degraded.

The results of a specific test data are illustrated in Fig. 6. CDSS demonstrates satisfactory performance without noticeable artifacts when using 64 (50%) channels. As the number of channels decreases, the majority of the structural information remains discernible in the results. Even with only 13 (10%) channels, CDSS can reconstruct the image effectively due to the increased measurement dimensions from 13 to 128 during the training stage, facilitated by the random masking strategy. In contrast, the conventional DAS method fails to distinguish the body's contour when the number of channels reduces to 38 (Fig. S2 in the supplementary materials). Comparatively, the result of CDSS with 10% channels exhibits superior image quality to DAS with 40% channels (Fig. S2 (c)).

We also evaluate the performances of CDSS and DAS with varying masking ratios from 90% to 50%. Fig. 7 depicts the trends of SSIM, PSNR and RMSE for DAS (blue) and CDSS (red) with different numbers of channels. CDSS consistently achieves an average SSIM score above 0.8. On the other hand, the SSIM, PSNR and RMSE of DAS are worse as the number of channels decreases. Even with only 13 (10%) channels, CDSS achieves satisfactory results (SSIM value of 0.876, PSNR value of 25.605 and RMSE value of 0.069). This SSIM value is higher than that of DAS with 128 channels (0.773, Table I). These findings highlight the effectiveness of dynamic random masking in achieving high reconstruction performance with extremely sparse view input by incorporating comprehensive information during training.

Based on the aforementioned results, we hypothesize that the random masking strategy can enhance the performance of supervised methods as well. To validate this hypothesis, we train a supervised post-processing Unet [65] using sparse input data with varying masking ratios. Specifically, during the training phase, we generate a mask randomly with a predetermined proportion (e.g., 50%) to select valid channels. We then use DAS method for reconstruction and feed the reconstructed image into Unet. Apart from that, the settings, including the dataset, remain consistent with those used in the supervised Unet. To facilitate a visual comparison, we also train a supervised Unet without the masking strategy, utilizing evenly distributed 16 channels.

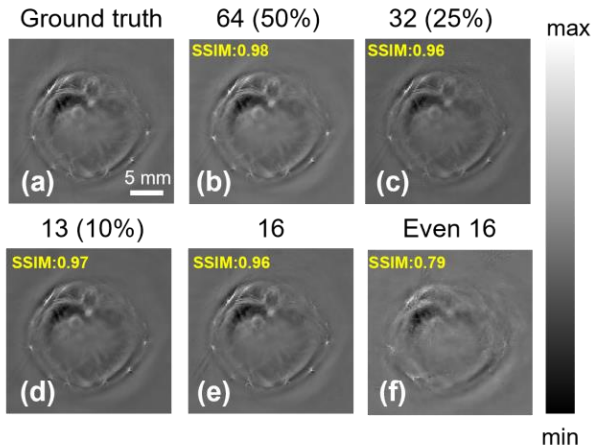

Fig. 8. The results of supervised Unet using different numbers of channels. (a) Ground truth reconstructed by DAS from 512 channels. The whole 128 channels are randomly masked with different ratios. The images reconstructed by supervised Unet from random (b) 64, (c) 32, (d) 13, and (e)16 channels. (f) The image reconstructed by supervised Unet without random masking from evenly distributed 16 channels.

In Fig. 8, the results of Unet with 64, 32, 16, and 13 channels are compared. The performance of Unet with random masking remains consistently high even as the number of channels decreases. The random masking technique facilitates an increase in the information available to the model during training, effectively expanding the information of input data through iterations. Fig. 8 (f) displays the outcome of Unet without the masking strategy, utilizing evenly distributed 16 channels. In comparison, Fig. 8 (e) exhibits more detailed information about the hepatic fault compared to Fig. 8 (f), indicating that the random masking strategy augments the measurement dimensions under sparse conditions. This observation further validates that Unet with random masking is capable of effectively incorporating the information equivalent to more channels during the training phase.

TABLE II.
QUANTITATIVE COMPARISONS OF THE SUPERVISED UNET WITH DIFFERENT NUMBERS OF CHANNEL'S (MEAN ± STANDARD DEVIATION)

|  | SSIM ↑ | PSNR ↑ | RMSE ↓ |
|---|---|---|---|
| 64 | 0.965±0.030 | 32.614±6.365 | 0.031±0.027 |
| 32 | 0.943±0.037 | 30.644±5.979 | 0.038±0.033 |
| 13 | 0.940±0.052 | 30.134±6.721 | 0.043±0.041 |
| 16 | 0.938±0.051 | 30.455±7.127 | 0.048±0.045 |
| even 16 | 0.774±0.057 | 22.256±4.287 | 0.082±0.043 |

Even 16 indicates the input image is reconstructed by supervised Unet from evenly distributed 16 channels without random masking.

Table II presents the quantitative results of supervised Unet using different numbers of channels. Across various sparse conditions, all the results exhibit similar performance when random masking is employed (SSIM > 0.9 and PSNR > 30). Notably, supervised Unet with evenly distributed 16 channels (0.774) does not surpass the performance of CDSS with 13 channels (0.832) in terms of SSIM. These findings highlight the scalability and potential of the random masking strategy to deliver effective reconstruction in sparse-view or limited-view scenarios, making it promising for implementation in different methods.

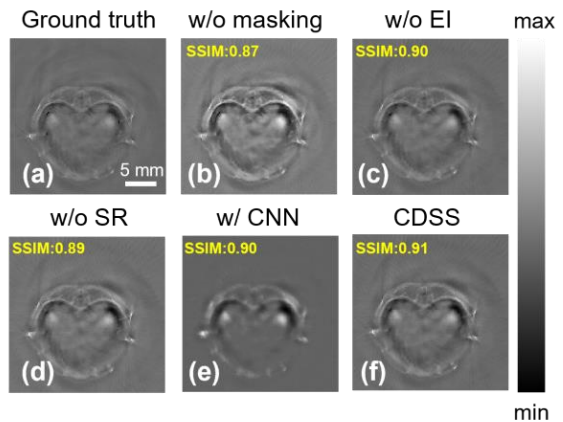

Fig. 9. The results of ablation studies. (a) Ground truth reconstructed by DAS from 512 channels. (b) The result of CDSS without random masking. (c) The result of CDSS without EI loss. (d) The result of CDSS without sparse regularization. (e) The result of CDSS using CNN model (Unet). (f) The result of CDSS.

### D. Ablation studies

In this section, our aim is to validate the impact of the following factors: (1) the random masking strategy, (2) the equivariant constraint, (3) the sparse regularization, and (4) the IFUnet model. To achieve this, we conduct the following ablation studies: (1) the proposed CDSS method utilizing 128 channels without masking, (2) the proposed CDSS method without the equivariant imaging loss, (3) the proposed CDSS method without sparse regularization, and (4) the proposed CDSS method with Unet model. All the experiments utilize data from 128 channels as labels. Additionally, for the experiments involving mask, the division is also made in a proportion corresponding to the 128 channels. It is worth noting that the masking ratio for all these experiments, except (1), is set to 50%.

Fig. 9 illustrates the results of the ablation experiments. Table III presents the quantitative results of the ablation experiments. Across all the experiments, our method consistently demonstrates improved performance in PACT image reconstruction.

TABLE III.
QUANTITATIVE COMPARISONS OF ABLATION STUDIES (MEAN ± STANDARD DEVIATION)

|  | SSIM ↑ | PSNR ↑ | RMSE ↓ |
|---|---|---|---|
| w/o masking | 0.903±0.032 | 28.659±4.215 | 0.042±0.022 |
| w/o EI | 0.901±0.121 | 30.535±7.042 | 0.038±0.041 |
| w/o SR | 0.892±0.105 | 30.742±7.040 | 0.042±0.045 |
| w/ CNN | 0.902±0.046 | 24.002±5.793 | 0.079±0.054 |
| CDSS | **0.917±0.049** | **32.422±6.591** | **0.033±0.033** |

## IV. DISCUSSIONS

In practical scenarios, obtaining complete measurement data for biomedical imaging reconstruction can be challenging and costly due to system constraints or scanning environments. This poses a significant problem for using DL in such scenarios: *Can the model learn to reconstruct high-quality images without access to ground truth images?* In this study, we have developed a novel cross-domain self-supervised approach to address this challenge and reconstruct PACT images using incomplete measurement data. The proposed approach offers valuable insights into overcoming the limitation of having imperfect training data and space for training.

Based on the physical principles, high-quality images can be obtained by utilizing measurements from dense detection channels. In computer vision, simple self-supervised methods have been employed in various models by randomly masking significant patches [70]. This concept inspired us to explore the idea of intentionally removing most semantic information during training, which could be more suitable for DL models. Consequently, we investigated the feasibility of masking channels to recover images from incomplete measurements. In the training phase, under the condition of limited quality images as labels, we further randomly divide the data into two complementary subsets. These partitioned subsets, along with the unmasked data, supervise each other to enhance the performance of reconstruction with fewer channels, while suppressing the different stripe artifacts simultaneously. Specifically, when the mask proportion reaches 10% and 90%, one of three parts only accesses dynamic 10% information of the data. However, with sufficient training iterations, this 10% information will gradually expand to all the channels. Namely, our method can still access 100% of the channel information during training, even if the mask ratio varies. This approach ensures that our model maintains its performance and accuracy, regardless of changes in the mask ratio. Previous work [41] demonstrated that Unet is unable to reconstruct high-quality images from measurements with a limited number of channels. However, by introducing an alterable masking strategy in each training batch to extend the information of the input, we observed that this strategy performs well even after masking most the channels, enabling effective mapping between images with 128 channels and images with 512 channels using the same supervised Unet model as in [41]. These findings indicate that the random masking strategy is scalable in image recovery and reconstruction.

In this study, we focused on demonstrating the feasibility of our proposed method in 2D image reconstruction, taking into account computational cost and the available data. However, it is important to note that the proposed CDSS method is not limited to 2D and can be potentially applied to reconstruct images in 3D scenes as well. The underlying principles and concepts of CDSS remain applicable in a 3D context. By leveraging the self-supervised learning approach and the equivariant imaging loss, CDSS can be adapted to address the challenges and complexities associated with 3D image reconstruction. In the future, further research and experimentation in the 3D scenes will help validate and expand the capabilities of CDSS for PACT reconstruction in volumetric data.

The present study identifies areas for improvement that will be addressed in future work. Specifically, we have observed that CDSS still exhibits some artifacts in the reconstructed images. These artifacts may be attributed to the fact that the only high-quality labels available during training are the images with 128 channels, leading to artifacts being retained in the resulting images. To address this issue, we intend to design modules that can effectively remove these artifacts and enhance the overall performance of our approach.

Another way for improvement is the incorporation of simultaneous supervision in both the signal and image domains, which has the potential to enhance the applicability and performance of the model. However, it is worth noting that the current performance of CDSS does not surpass that of the supervised method. Although the equivariant constraint expands the space of limited measurements, the information obtained by CDSS is still not on par with that reconstructed using 512 measurement channels. To address this limitation, introducing additional physical constraints could alleviate the problem and help prevent instability that may arise from having more regularization parameters. By incorporating more physical constraints, we can aim to improve the reconstruction quality and overall performance of CDSS.

Finally, the pseudo-inverse matrix was established based on the TOF in this work. However, certain acoustic properties, such as attenuation, were not considered, leading to errors in the directly reconstructed image. To improve the accuracy of loss computation and enhance the quality of the directly reconstructed image, it would be beneficial to incorporate a space-variant point spread function (PSF) based model. By accounting for additional acoustic properties, such as attenuation, the reconstruction process can be more accurate, resulting in improved image quality. This is an avenue for future research to enhance the accuracy and fidelity of the directly reconstructed images in PACT.

## V. Conclusion

In this study, we introduce a novel cross-domain self-supervised learning approach for PACT image reconstruction, which eliminates the need for full sampled measurements. By formulating the reconstruction problem in a similar objective target as the model-based methods. CDSS effectively leverages the available information in the measurement data to achieve accurate and high-quality reconstructions. By randomly masking measurement channels, we reconstruct the image using fewer PA data from the unmasked channels, exploiting the redundancy in the measurement data. This approach enables self-supervision in both the raw data and image spaces. Additionally, we leverage the equivariance of PACT for arbitrary rotations to introduce a novel equivariant imaging loss, without requiring additional distributions of underlying data. Experimental results on *in-vivo* mice datasets with varying numbers of channels demonstrate that our proposed CDSS approach can achieve highly accurate image reconstruction with limited measurement channels. Surprisingly, even with extremely few signals (e.g., 13 channels), CDSS outperforms supervised CNN methods. Furthermore, compared to conventional model-based methods, CDSS significantly reduces the reconstruction time. Our study also introduces a new end-to-end model for PACT image reconstruction. While self-supervised schemes may not fully surpass supervised methods, our CDSS framework demonstrates the feasibility of a self-supervised perspective in medical image reconstruction when combined with physical models. These findings inspire new possibilities for enhancing medical image reconstruction by incorporating self-supervision and leveraging the underlying physical principles.